\documentclass[a4paper,conference,final]{IEEEtran}

\usepackage{siunitx}
\DeclareSIUnit \dBm {dBm}
\DeclareSIUnit \dB {dB} 
\DeclareSIUnit \dBi {dBi} 
\DeclareSIUnit \Kbps {Kbps}
\DeclareSIUnit \Mbps {Mbps}
\DeclareSIUnit \Gbps {Gbps}
\DeclareSIUnit \kBps {kBps}
\DeclareSIUnit \MBps {MBps}
\DeclareSIUnit \GBps {GBps}

\usepackage{cite}      
\usepackage[caption=false, font=footnotesize]{subfig}
\usepackage[pdftex]{graphicx}
\usepackage{url}       
\usepackage{amsfonts}
\usepackage{amsmath}
\usepackage{times}
\usepackage{algpseudocode}
\usepackage{fourier}
\usepackage{algorithm}
\usepackage{enumerate}
\usepackage{multirow}
\usepackage{tikz}
\usepackage{array}
\usepackage{gensymb}
\newcolumntype{P}[1]{>{\centering\arraybackslash}p{#1}}
\newcolumntype{M}[1]{>{\centering\arraybackslash}m{#1}}
\DeclareMathOperator*{\argmax}{arg\>max}
\newcommand*{\argmaxl}{\argmax\limits}
\bstctlcite{IEEEexample:BSTcontrol}

\begin{document}
\title{MmWave System for Future ITS: A MAC-layer Approach for V2X Beam Steering}
\author{\IEEEauthorblockN{Ioannis Mavromatis, Andrea Tassi, Robert J. Piechocki, and Andrew Nix}
  \IEEEauthorblockA{Department of Electrical and Electronic Engineering, University of Bristol, UK \\ Emails: \{ioan.mavromatis, a.tassi, r.j.piechocki, andy.nix\}@bristol.ac.uk}
}

\maketitle

\begin{abstract}
Millimetre Waves (mmWave) systems have the potential of enabling multi-gigabit-per-second communications in future Intelligent Transportation Systems (ITSs). Unfortunately, because of the increased vehicular mobility, they require frequent antenna beam realignments - thus significantly increasing the in-band Beamforming (BF) overhead. In this paper, we propose Smart Motion-prediction Beam Alignment (SAMBA), a MAC-layer algorithm that exploits the information broadcast via DSRC beacons by all vehicles. Based on these information, overhead-free BF is achieved by estimating the position of the vehicle and predicting its motion. Moreover, adapting the beamwidth with respect to the estimated position can further enhance the performance. Our investigation shows that SAMBA outperforms the IEEE 802.11ad BF strategy, increasing the data rate by more than twice for sparse vehicle density while enhancing the network throughput proportionally to the number of vehicles. Furthermore, SAMBA was proven to be more efficient compared to legacy BF algorithm under highly dynamic vehicular environments and hence, a viable solution for future ITS services.
\end{abstract}

\begin{IEEEkeywords}
Connected Autonomous Vehicles, mmWave, Beamforming, Heterogeneity, MAC layer, Vehicle-to-Everything Communications.
\end{IEEEkeywords}

\vspace{-0.1cm}
\section{Introduction}
Connected and Autonomous Vehicles (CAVs) will act as key entities for \emph{Next-Generation Intelligent Transportation System (ITS) applications and services}. Vehicles being gradually equipped with more sensors, will have the potential of enhancing transportation safety and reaching full autonomy~\cite{cavs}. Sensory data distributed to the surrounding vehicles can be used to better understand the traffic conditions and improve navigation quality. The same data, shared with the infrastructure network, can exploit cloud computing capabilities for efficient resource management~\cite{cloud_computing}, extend the network scalability and provide access to essential applications (e.g., spectrum sharing, dissemination, etc.)~\cite{vehicle_cloud}. 

The next-generation automotive applications will require \emph{gigabit-per-second data rates} and \emph{tactile-like end-to-end delays}, introducing very strict Quality of Service (QoS) requirements~\cite{qos_req}. These QoS constraints cannot be adequately supported with Dedicated Short Range Communications (DSRC), as
IEEE 802.11p/DSRC can achieve up to \SI{27}{\Mbps} with modest delay performance~\cite{11p_delay}. Alternative, Millimetre Wave (mmWave) communications can effectively fulfil these requirements. However, mmWave propagation characteristics, combined with the increased mobility in vehicular environments, lead to performance degradation due to Doppler shifts and frequent misalignments.

As a solution, we propose Smart Motion-prediction Beam Alignment (SAMBA) algorithm. Our algorithm, operates in a heterogeneous manner combining DSRC and mmWave Radio Access Technologies (RATs) and leveraging from the position and the motion information broadcast from a CAV, it enhances the Beamforming (BF) process. SAMBA, reducing the beam misalignments under the highly dynamic vehicular environments, manages to improve the mmWave system performance. Providing solutions to problems arising from other BF techniques, can make the adoption of mmWaves for vehicular communications easier.


Referring to the existing mmWave standards, the BF procedure requires a bidirectional frame exchange, operating with quasi-omnidirectional patterns in an beam-sweeping manner. However, higher vehicle velocities introduce an increased Doppler spread and traditional BF processes fail~\cite{coherence_time}. For lower speeds and Line-of-Sight (LOS) links, the Doppler shift can be corrected with frequency offset correction techniques. However, the beam-sweeping increases dramatically the delay. According to~\cite{beamforming_delay}, the response time required from the chip to change the phase and the gain of a phased-array antenna is roughly $\simeq$\SI{50}{\nano\second}, proving that the BF delay is related with the number of frames exchanged.



Mobile systems require frequent beam steering. This leads to significant in-band overhead. Leveraging from the idea of heterogeneity, zero in-band overhead can be achieved. Authors in~\cite{eyes_closed}, train the antenna beams by passively overhearing frames in the legacy band of \SI{2.4}/\SI{5}{\giga\hertz} and estimating the Angle-of-Arrival (AoA). Though, in dense urban environments, AoA is not accurately estimated due to the multipath effects. Feedback information from a vehicle, sent over DSRC links in the form of beacons, can be facilitated to overcome that. In~\cite{beam_design}, a vehicle transmitting its initial position and speed, provides feedback for the infrastructure-side BF. However, position errors were not taken into account, vehicle speed was constant and no complex manoeuvres were considered limiting the utilisation of the algorithm on a straight-road scenario. Similarly, SAMBA can enhance the performance by fusing the position, motion and velocity data from a vehicle, making it able to operate on wider scale complex road networks.


A certain level of accuracy should be achieved when basing the system performance on a node-localisation system.
The most inaccurate measurement is related to the position. Commonly acquired via Global Positioning System (GPS), it is affected by additive errors. Especially in urban environments, urban street canyon effects can be observed due to the height of the buildings - thus reducing the position accuracy. However, the accuracy can be significantly improved by fusing the position data with the motion information of a vehicle, achieving centimetre-accuracy in urban environments~\cite{cent_accuracy}.
Other approaches for lane positioning based on sensor networks and inter-vehicle communications can be found in~\cite{lane_positioning} and~\cite{markov_based_positioning}, presenting highly accurate results. CAVs equipped with numerous sensors, RATs, and increased processing power will be able to acquire and feedback their accurate position. The above serve as a proof of the capacity required for the level of the position accuracy necessary for our algorithm.


This paper is organised as follows. In Section~\ref{sec:system_model}, the system model, the integration of IEEE 802.11ad into ITS applications, and the problem motivation are introduced. The proposed algorithm, the models and the assumptions are presented in Section~\ref{sec:smart_algorithm}. Section~\ref{sec:results}, presents the simulation framework and SAMBA performance is compared with the IEEE 802.11ad BF. Finally, our work is summarised in Section~\ref{sec:conclusions} and ideas for future research are mentioned.

\section{System Model and Problem Motivation}\label{sec:system_model}

Road Side Units (RSUs) are fitted along the side of the road or at essential locations (e.g., intersections), usually mounted on street light poles or traffic lights, \SI{6}-\SI{10}{\meter} higher than the level of the vehicles. In such manner, blockage from other vehicles can be avoided and LOS links can be established, as analysed in~\cite{beam_design}. Moving vehicles need to frequently realign their antenna beams either with the serving RSU or with another vehicle. In this work we will focus on a Vehicle-to-Infrastructure (V2I) scenario.

\subsection{Traditional Beamforming for Vehicular Networks} 
Consider a scenario that utilises only mmWaves RAT and two devices, a RSU and a vehicle. The distance between them, combined with the mmWaves propagation characteristics, imply the necessity for BF to maximise the data rate. Referring to IEEE 802.11ad~\cite{standard}, the dominant standard for mmWaves, the BF training requires a bidirectional frame exchange.  The MAC layer of IEEE 802.11ad introduces the concept of \emph{virtual antenna sectors}, discretizing the azimuth plane based on the antenna beamwidth~\cite{standard}.

The BF training is split in two different intervals. During \emph{Sector Level Sweep (SLS)} interval, the RSU beams are trained transmitting directional frames on each sector in a sweeping manner. The vehicle, listening quasi-omnidirectionally, transmits feedback information for each frame and the best RSU sector is chosen based on the highest Signal-to-Noise Ratio (SNR). Later, during \emph{Beam Refinement Protocol (BRP)}, the vehicle chooses its best antenna sector, similarly as before, establishing a directional link. Finally, the beams are further refined. Testing multiple configuration on the already established link, the quasi-omnidirectional imperfections are avoided and the antenna array configurations can be fine-tuned maximising the achievable data rate.

When the vehicle number increases, the responding interval within SLS is slotted (\emph{Association Beamforming Training (A-BFT)}). Each vehicle is allocated one timeslot based on a uniform random distribution with values $U(4,8)$~\cite{standard}. During this slot, the vehicle transmits its feedback information, further refining its beams later during BRP.

\begin{figure}[!tbp]
  \centering
    \subfloat[Collision probability during one A-BFT for different number of vehicles.]{	\includegraphics[width=1\columnwidth]{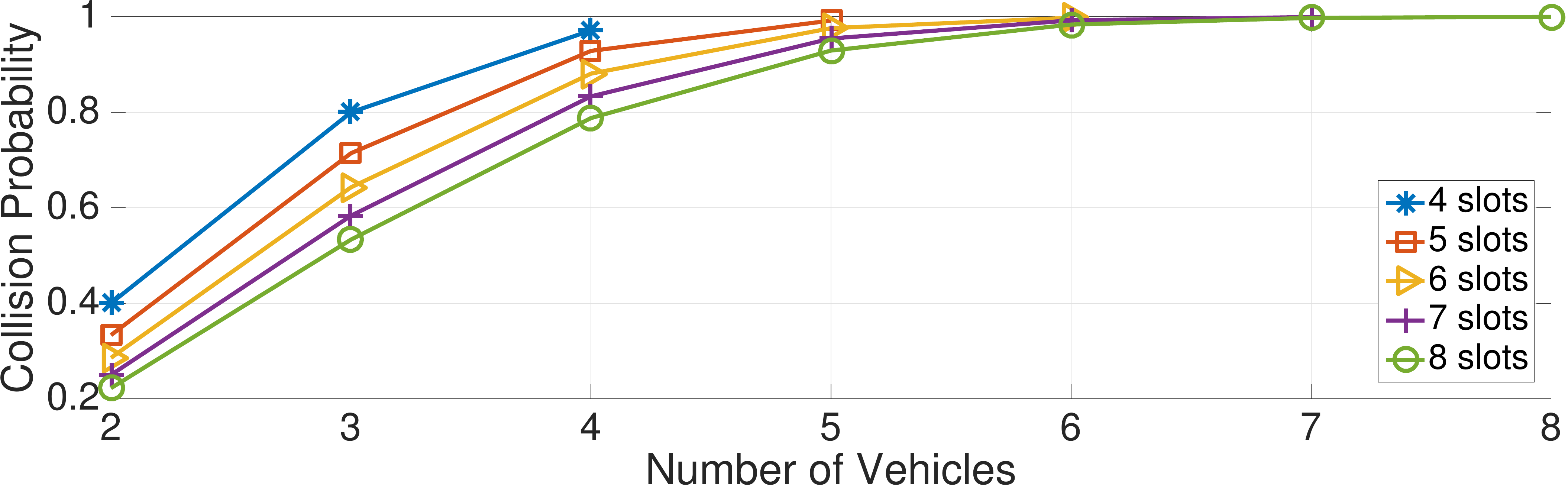}
    \label{fig:coll_prob}}
  \hfill
  \subfloat[Average delay introduced from legacy BF training every Beacon Interval. 16 virtual sectors were assumed for each antenna.]{\includegraphics[width=1\columnwidth]{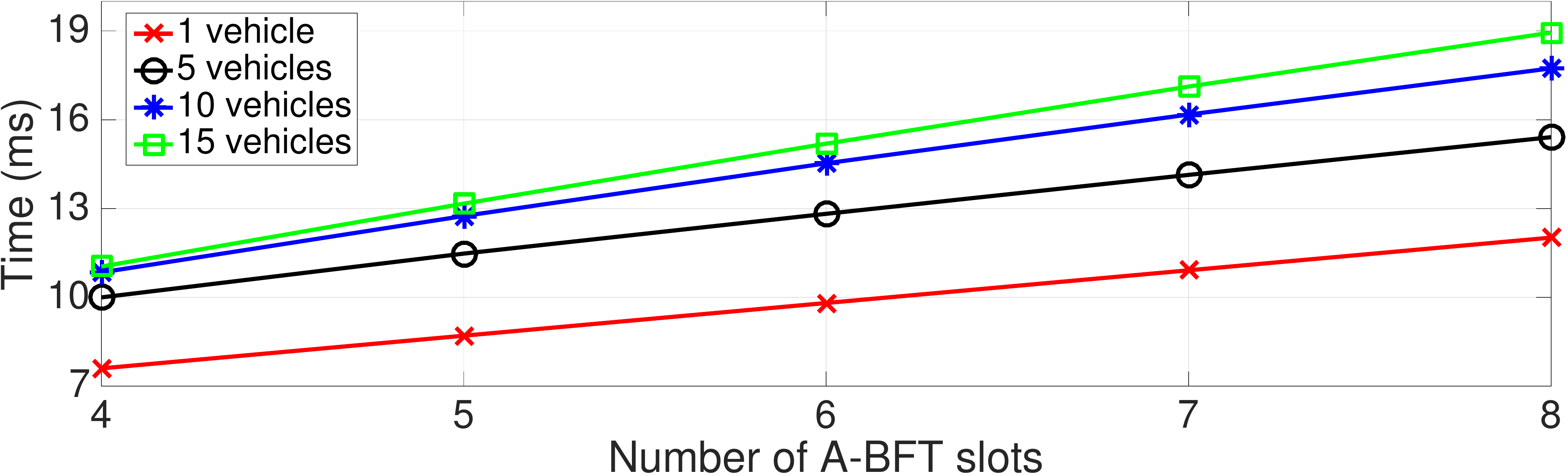}
    \label{fig:timemisspent}}
    
    \caption{The number of non-trained beams (due to collisions) and the BF delay, can severely impact the performance of vehicular communications.}
	\label{fig:coll_timemisspent}
\end{figure}

\subsection{Challenges of Legacy BF Strategy for Future ITSs}\label{subsec:11ad_vanet}
The above BF procedure is performed every one \emph{Beacon Interval (BI)}~\cite{standard}. BI length is limited to \SI{1000}{\milli\second} and can be optimised for each environment. For example, indoor environments with zero or low mobility require a value within the range of \SI{100}{\milli\second}. Longer intervals reduce the management frame transmission rates and increase throughput, however the system becomes intolerant to Doppler Shift. Vehicular communications require shorter intervals ($<$\SI{30}{\milli\second}), to avoid severe performance degradation due to beam misalignments from the increased mobility. 


The slotted A-BFT with the random slot allocation leads to collisions. Also, the predefined number of slots is insufficient for urban scenarios, as more than eight vehicles can convene within a RSU coverage region. The \emph{probability of collision} within a A-BFT slot can be defined as follows:
\begin{equation}
P_{\mathrm{col}} =  1 - \dfrac{x! \, \left( x - 1 \right)!}{ \left( x - v \right)! \, \left( x + v -1 \right)!}
\end{equation}
where $x$ is the number of slots and $v$ is the number of vehicles. The probability of avoiding a collision is the ratio of the combinations when only one vehicle is allocated per slot $\binom xv$, over the number of vehicles allocated to a number of slots $\binom {x+v-1}v$. $P_{\mathrm{col}}$ is the complement of the above.


The collision probability during A-BFT dramatically increases as the number of vehicles is increased (Fig.~\ref{fig:coll_prob}) - thus, significantly reducing the trained antenna beams. Furthermore, a full-circle sector sweeping is not always necessary as some sectors might point towards a direction with no vehicles. Based on that, an overhead analysis for IEEE 802.11ad BF technique was introduced in our previous work~\cite{mywork}, deriving formulas to approximate the delay introduced for $N$ number of vehicles.   The results are shown in Fig.~\ref{fig:timemisspent}. The aforesaid prove that legacy BF is incapable of fulfilling the strict QoS requirements of Next-Generation ITSs for tactile-like end-to-end delays ($<$\SI{10}{\milli\second}) and gigabit-per-second throughput.

\begin{figure}[t]     
\centering
    \includegraphics[width=0.9\columnwidth]{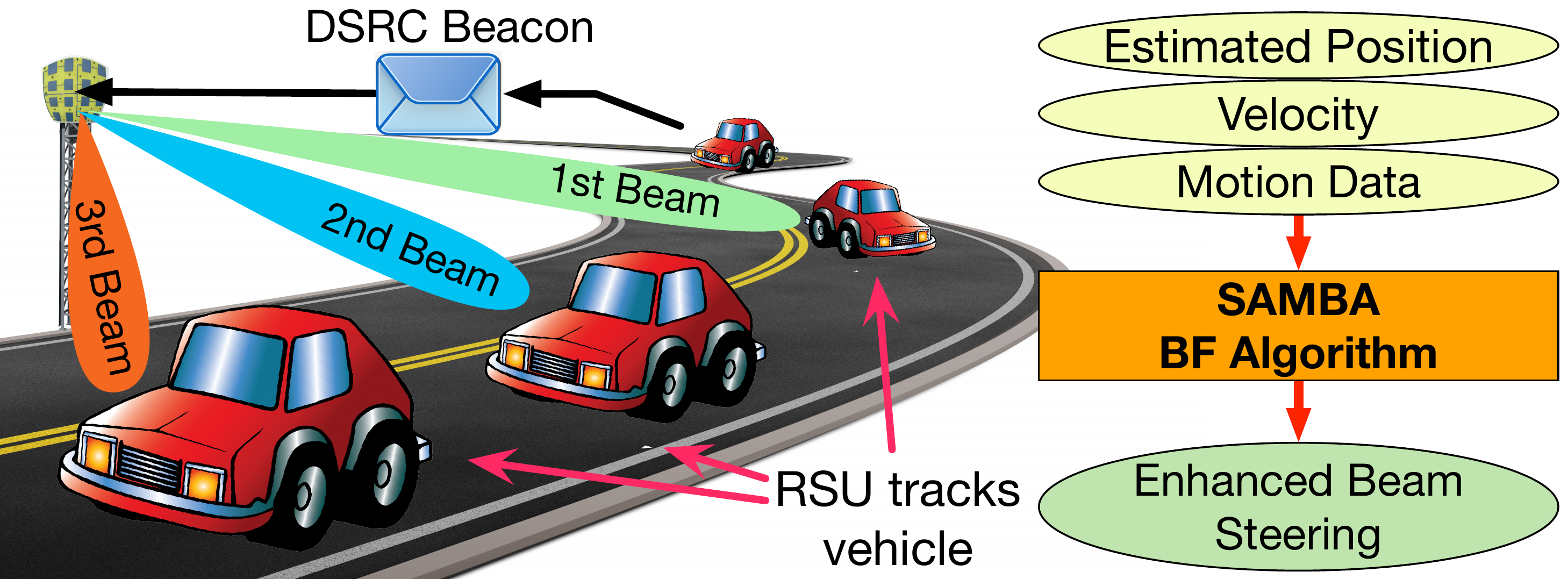}
    \caption{SAMBA System design: Position, velocity and motion information encapsulated in DSRC beacons are used for smart BF. RSUs predict the motion of the vehicle and its position and align their beams accordingly.}
    \label{fig:beam_alignment}
\end{figure}

\section{SAMBA: Enhanced Beamforming for V2I Links}\label{sec:smart_algorithm}
To solve the previously mentioned problems, we proposed the SAMBA algorithm. Leveraging from the position, the velocity and the motion information broadcast, SAMBA can provide an overhead-free BF, reduce the association delays, minimise the beam misalignments and enhance mmWave performance. SAMBA operates as shown in Fig.~\ref{fig:beam_alignment}.

On the infrastructure side, and as shown in Alg.~\ref{alg:bf_agl_rsu}, SAMBA algorithm considers a road network with $N$ number of vehicles, where $N\geq 1$. The information received from a vehicle is used to decide whether the vehicle has moved, compared to the previously stored position. With respect to all the positions, the RSUs decide their serving vehicles, i.e., each vehicle is served by its closest RSU. When an updated position is received, the serving RSU aligns its beam accordingly. Later, the RSUs can effectively track the movement of each vehicle predicting its motion and position. The update interval for SAMBA was predefined at \SI{30}{\milli\second}. By that, a comparable BI with IEEE 802.11ad is used and increased Doppler shift can be avoided.

On the vehicle side, the vehicles transmit DSRC beacons to all RSUs in range encapsulating their velocity, their motion data (based on the \emph{vehicle motion dynamics}) and their estimated position. Beacons are broadcast every \SI{100}{\milli\second} (DSRC beacon interval) and the acquired information is updated periodically. All CAVs, as smart entities of an ITS, can a priory know the positions of the RSUs. To that extent, each vehicle aligns its beam towards the closest RSU.


When more than one vehicles are within the coverage region of a RSU, a dynamic channel time allocation access mechanism is used, that implements a polling based channel access, similar to the one of IEEE 802.11ad~\cite{standard}. Using the same mechanism for both approaches, it can be ensured that the delays introduced from the resource allocation scheme are negligible for our results.



\begin{algorithm}[t]

\caption{SAMBA Algorithm: Infrastructure Side}
\label{alg:bf_agl_rsu}
{\footnotesize\begin{algorithmic}[1]
    \Require Vehicles encapsulate position, motion and velocity in beacons
    \Ensure $RSU_{\mathrm{n}}$ has not changed after every update interval.
	\While{$N$ number of Vehicles within the network range ($N\geq 1$)}
      \If{New beacon received}
              \State Find $RSU_{\mathrm{n}}$ for each vehicle \Comment{$RSU_{\mathrm{n}}\rightarrow Closest~RSU$}
              \If{$Received_{\mathrm{pos}}\neq P_{\mathrm{pos}}$}
                  \State Beamforming: Align $RSU_{\mathrm{n}}$ beam based on $Received_{\mathrm{pos}}$
              \Else
                  \State Predict current position of vehicle $P_{\mathrm{pos}}$
                  \State Beamforming: Align $RSU_{\mathrm{n}}$ beam based on $P_{\mathrm{pos}}$
              \EndIf
      \Else
              \Repeat~every Update Interval       \Comment{\SI{30}{\milli\second}}
                  \State Predict current position of vehicle $P_{\mathrm{pos}}$
                  \State Beamforming: Align $RSU_{\mathrm{n}}$ beam based on $P_{\mathrm{pos}}$
              \Until{New beacon is received}
      \EndIf
    \EndWhile
\end{algorithmic}}
\end{algorithm}

\subsection{Mobility Model and Position-based Beam Alignment}\label{subsec:mob_model}
An urban scenario can be accurately represented by the synchronised flow traffic model~\cite{sync_flow}. It represents a continuous traffic flow, with no significant stoppages, where vehicles perform random manoeuvres (braking/accelerating, changing lanes) and tend to synchronise their movement. Velocity varies over time and is averaged around a mean value, following a Normal distribution, i.e., $s\sim \mathcal{N}(s_{\mathrm{avg}},2)$. Velocity error can be easily corrected by means of data fusion techniques and was not considered in this model. 

The estimated vehicle position is affected by an additive error. A typical mean error for GPS is about \SI{3}{\meter} with a standard deviation of roughly \SI{1}{\meter}~\cite{gps_accuracy}.  As discussed though, increased accuracy can currently be achieved by fusing CAVs sensory data, even under urban environments~\cite{cent_accuracy}. The estimated position is given as $E_{\mathrm{pos}} = R_{\mathrm{pos}} + e_{\mathrm{pos}}$, where $R_{\mathrm{pos}}$ is the real position of the vehicle and $e_{\mathrm{pos}}\sim\log\mathcal{N}(\mu,\sigma^2_s)$ is the log-Normal error. Terms $\mu$ and $\sigma$ are the non-logarithmetised values for the mean and standard deviation of the log-Normal distribution. Knowing $E_{\mathrm{pos}}$, a RSU can steer its beam accordingly by calculating the angle $k$ (Fig.~\ref{fig:one_beam}) with respect to the reference plane (from the trigonometric properties of right-angled triangles) .


\subsection{Vehicle Motion Dynamics and Motion-Prediction}\label{subsec:veh_dyn}
CAVs equipped with Inertial Measurement Unit (IMU) sensors (e.g., magnetometers, accelerometers, gyroscopes), will be able to measure the motion changes of a vehicle. The acquired sensory data, can be combined using data fusion algorithms. Their output is the angular velocity of the vehicle, measured as \SI{}{\radian\per\second}, in three different axis (\emph{yaw $\omega_\mathrm{y}$, pitch $\omega_\mathrm{p}$, roll $\omega_\mathrm{r}$}). Sensory data errors are within the range of $0.2\degree-1\degree$~\cite{cent_accuracy} and does not introduce significant errors in our algorithm, therefore they were considered as negligible.

\begin{figure}[!tbp]
  \centering
    \subfloat[Beam steered based on the position after the additive error.]{	\includegraphics[width=0.47\columnwidth]{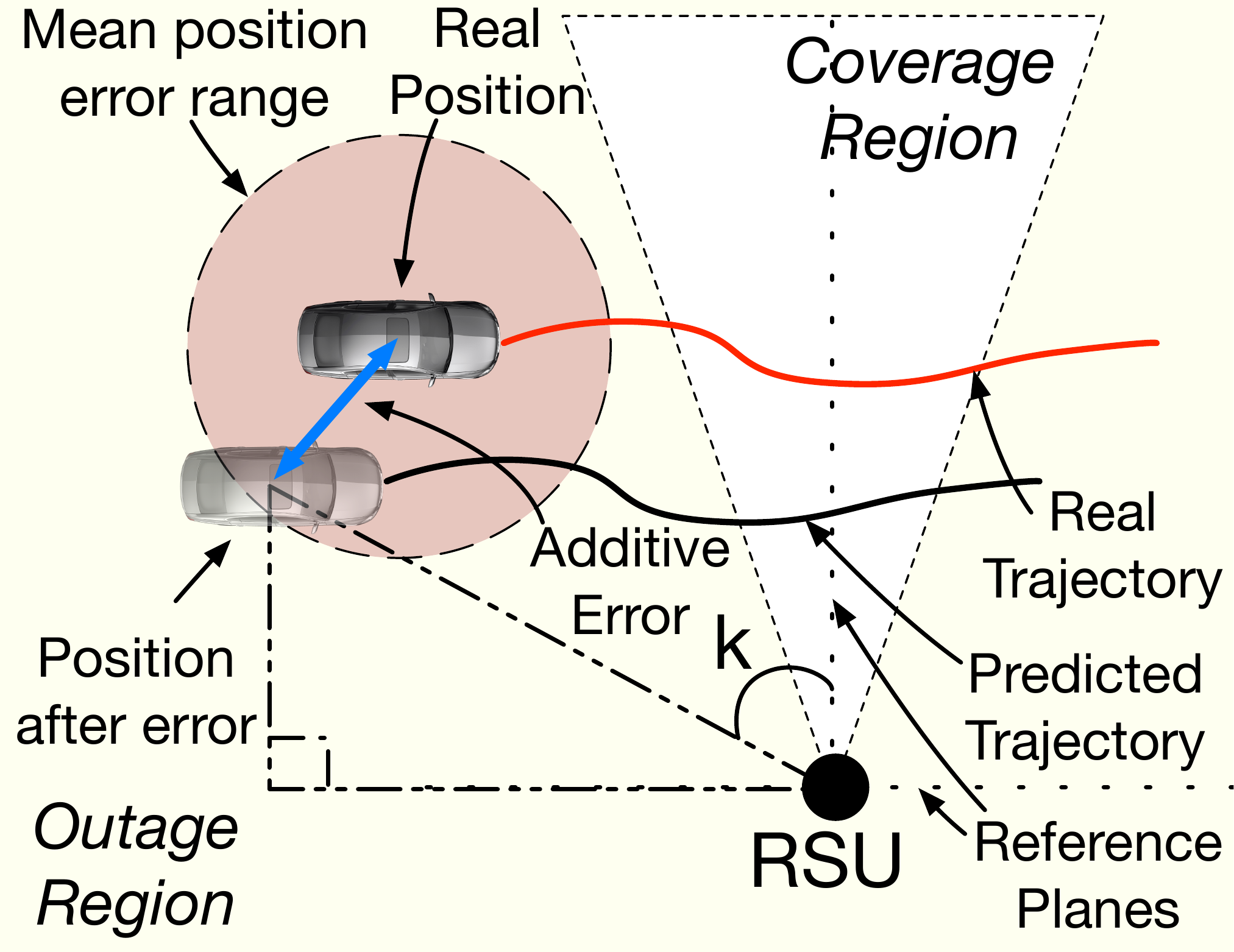}
    \label{fig:one_beam}}
  \hfill
  \subfloat[Vehicle moving on the perimeter of a circle (2D representation).]{\includegraphics[width=0.47\columnwidth]{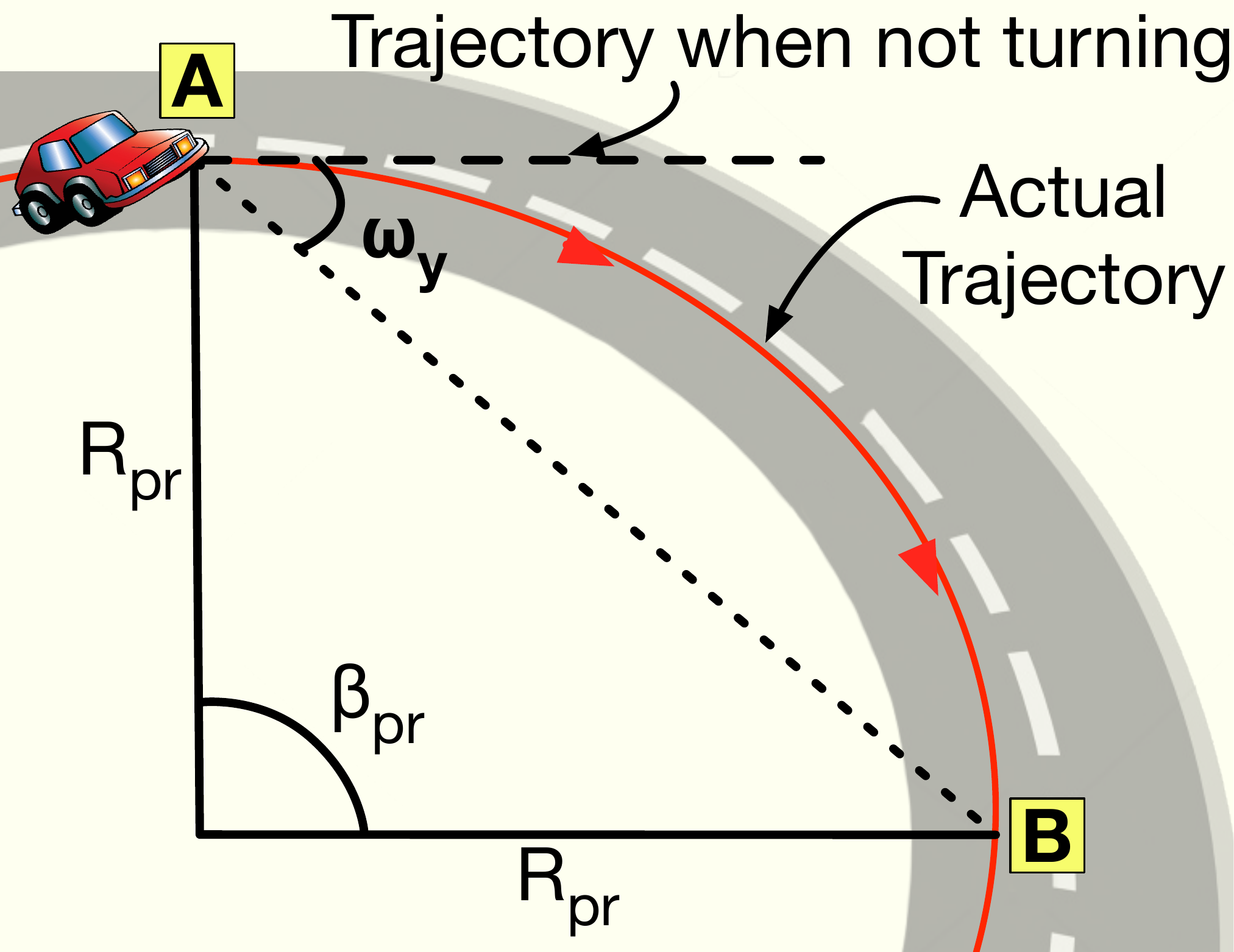}
    \label{fig:veh_motion}}
    
    \caption{a) Beam steering, and how the position error introduces a different to the trajectories. b) Motion of vehicle within a time interval.}
	\label{fig:motion_update}
\end{figure}


Consider a constant angular speed. A vehicle in motion follows the surface of a sphere (when observed within one time interval). Vehicles change significantly their direction  on the vertical axis, as they are changing their direction on the road plane. So, in this work, vehicles and RSUs will be considered as 2D objects, positioned on a plane. In this case, a vehicle follows the perimeter of a circle and its motion can be predicted based on $\omega_\mathrm{y}$, $E_{\mathrm{pos}}$, and $s_{\mathrm{avg}}$. 

With respect to Fig.~\ref{fig:veh_motion}, a vehicle moving from $A$ to $B$, will drive a distance of $l\widearc{AB}_{\mathrm{pr}}$. The distance travelled and its angle $\beta_{\mathrm{pr}}$, can be defined as follows:
\begin{equation}\label{eq:eq5}
\beta_{\mathrm{pr}} = m\widearc{AB} = 2 \, \omega_\mathrm{y} \, t_{\mathrm{pr}}
\qquad
l\widearc{AB}_{\mathrm{pr}} = s \, t_{\mathrm{pr}}
\end{equation}
where $t_{\mathrm{pr}}$ is the time elapsed from the latest beacon and $s$ is the velocity of the vehicle. Based on the circle properties and using \eqref{eq:eq5}, the radius of the circle $R_{\mathrm{pr}}$, and the distance $\overline{AB}_{\mathrm{pr}}$ between points $A$ and $B$, are given as follows:
\begin{equation}\label{eq:eq6}
R_{\mathrm{pr}} = \dfrac{ l\widearc{AB}_{\mathrm{pr}}}{\pi \, \omega_\mathrm{y} \, t_{\mathrm{pr}}}
\qquad
\overline{AB}_{\mathrm{pr}} = 2 \, R_{\mathrm{pr}} \, \sin(\omega_\mathrm{y} \, t_{\mathrm{pr}})
\end{equation}
Finally, the predicted position $P_{\mathrm{pos}}$\footnote{The model can be extended to a 3D scenario, by modifying~\eqref{eq:eq5},~\eqref{eq:eq6} and~\eqref{eq:eq7} to fit a spherical object.} is calculated as:
\begin{equation}\label{eq:eq7}
P_{\mathrm{pos}}(x,y) = \begin{cases}
	P_{\mathrm{pos}}(x) = E_{\mathrm{pos}}(x) + \overline{AB}_{\mathrm{pr}} \, \sin(\beta_{\mathrm{pr}}) \\
	P_{\mathrm{pos}}(y) = E_{\mathrm{pos}}(y) + \overline{AB}_{\mathrm{pr}} \, \cos(\beta_{\mathrm{pr}})
	\end{cases}
\end{equation}


\subsection{SNR, Antenna Gain, and Link Budget Analysis}\label{sub:link_budget}
Aligned beams imply higher SNR. The wireless standards define sensitivity thresholds for each Modulation and Coding Scheme (MCS). Each MCS can be associated with the SNR to optimise the data rate. The SNR is expressed as the ratio between the \emph{received power}  over the \emph{noise power}, i.e., $\mathrm{SNR} = P_{\mathrm{RX}}/P_{\mathrm{noise}}$, and is affected by the antenna gains. 

The antenna gain is related to its beamwidth. In this work, an ideal beam is assumed with uniform gain and no sidelobes. The directivity of an antenna is equal to $D = 4\pi/\Omega_{\mathrm{A}}$, where $\Omega_{\mathrm{A}}$ is the beam solid angle~\cite{balanis}. For our model $\Omega_A\approx \theta_{1} \, \theta_{2} $, where $\theta_{1}$ and $\theta_{2}$ are the half-power (\SI{-3}{\dB}) beamwidths of the elevation and azimuthal polarisation planes respectively. The antenna gain $G$ is proportional to the antenna efficiency $\eta$ and its directivity, and is given as $G = \eta \, D$~\cite{balanis}. For an ideal antenna, the efficiency is equal to $100\%$ and $\theta_{1}=\theta_{2}$, so the beamwidths in both polarisation planes become equal to the antenna beamwidth $\theta$. From all the above, $G$ is given as follows:
\begin{equation}
G(\theta) \simeq 4 \, \pi/\theta^2
\end{equation} 

The $P_{\mathrm{RX}}$ is equal to~\cite{prediction_model}:
\begin{equation}
P_{\mathrm{RX}} = P_{\mathrm{TX}} + G_{\mathrm{RX}} + G_{\mathrm{TX}} - PL
\end{equation}
where $P_{\mathrm{TX}}$ is the transmission power and $G_{\mathrm{TX}}$-$G_{\mathrm{RX}}$ are the antenna gains for the transmitter and the receiver respectively. In this work, equal beamwidth and thus equal gain was considered for both antennas. The $PL$ is the \emph{path-loss component} and can be calculated as:
\begin{equation}
PL = 10 \, n \, \log_{10}d + C_{\mathrm{att}} + A_{\mathrm{att}} + R_{\mathrm{att}} + S_{\mathrm{f}} 
\end{equation}
where $n$ is the path-loss exponent and $d$ is the separation distance between the RSU and the vehicle. $A_{\mathrm{att}}$ and $R_{\mathrm{att}}$ are the average atmospheric and rain attenuation, respectively. $C_{\mathrm{att}}$ represents the channel attenuation for a mmWave LOS link at \SI{60}{\giga\hertz} in an urban environment, measured at \SI{20}{\meter}~\cite{prediction_model}. Finally, $S_{\mathrm{f}}$ is a random shadow fading of the channel and it follows a log-Normal distribution $S_{\mathrm{f}}\sim\log\mathcal{N} (0,\sigma^2_{SF})$ with $\sigma = 5.8$~\cite{sigma_sf}.  

Finally, $P_{\mathrm{noise}}$ is as follows:
\begin{equation}
P_{\mathrm{noise}} = N_{\mathrm{fl}}+10\log_{10}B+N_{\mathrm{fig}}
\end{equation}
where $N_{\mathrm{fl}}$ is the noise floor value, $N_{\mathrm{fig}}$ is the noise figure, and $B$ is the channel bandwidth. 
For a given SNR, the most appropriate MCS can be found comparing the sensitivity thresholds with the SNR, and choosing one able to compensate with the noise level. For this work, seven MCSs were used based on IEEE 802.11ad standard~\cite{standard}.

\subsection{Beamwidth Adaptation to Maximise the Data Rate}

Consider a scenario where a vehicle travels on a road, approaching the RSU from distance, passing by and fending off until it is outside of the coverage region. It is observed that the beam covers a much wider area at its edge. A wide beam implies a wide beamwidth and consequently low antenna gains and SNR. To maximise the performance, an optimal solution is narrower beams away from the RSU, increasing the beamwidth as the vehicle gets closer.

As SAMBA relies on the position information, the error introduced will lead to misalignments for very narrow beams. From geometry is know that the incentre is the centre of triangular area and is equally spaced from the beam edges. To that extent, adapting $\theta$ accordingly, the vehicle can be positioned at the centre of the beam. 

Centring the vehicle with respect to the beam edges will maximise the data rate and minimise the misalignments compensating with the random errors.  To do so, the incentre point is assumed to be $E_\mathrm{pos}$ and $d_\mathrm{in}$ is the distance from the RSU. To maximise the data rate, the highest MCS should be used, i.e., the sensitivity threshold $U$ of the highest MCS is greater than the SNR. The above can be expressed as:
\begin{equation}\label{eq:argmax}
\begin{array}{rrcl}
\hat{\theta} & = & \argmaxl_{\theta} & \Big\{ D_\mathrm{i}\,
(\,d_\mathrm{in},\,\theta_\mathrm{i}\,)\,\Big\}
\\
& & \textrm{s.t.} & i = 1,2,...,N,~U \geq \gamma_\mathrm{i},~\theta_\mathrm{i}>0,~~~\forall i
\end{array}
\end{equation}
where $\hat{\theta}$ is the adapted beamwidth, $D_\mathrm{i}$ is the MCS data rate, $\gamma_\mathrm{i}$ is the given SNR for a position and, finally, $N$ is the number of required beam realignments. 



\vspace{-0.1cm}
\section{Performance Evaluation}\label{sec:results}
\begin{table}[t]
\renewcommand{\arraystretch}{1.07}
\centering
    \caption{List of Simulation Parameters.}
    \begin{tabular}{rl|rl}

    \textbf{Parameter}  &           & \textbf{Value}    & \\ \hline \hline
    Carrier Frequency   & $f_{\mathrm{c}}$     & \SI{60}           & \SI{}{\giga\hertz}  \\ 
    Bandwidth 		    & $B$       & \SI{2.16}         & \SI{}{\giga\hertz} \\ 
    Path-Loss Exponent  & $n$       & 2.66              & ~\cite{path_loss}      \\ 
    Atmospheric Attenuation  & $A_{\mathrm{att}}$  & \SI{15}   & \SI{}{\dB\per\kilo\meter}      \\ 
    Rain Attenuation    & $R_{\mathrm{att}}$       & \SI{25}  & \SI{}{\dB\per\kilo\meter} (in the UK)     \\ 
    Channel Attenuation & $C_{\mathrm{att}}$  & \SI{70}       & \SI{}{\dB}~\cite{prediction_model}  \\
    Transmission power  & $P_{\mathrm{TX}}$  & \SI{10}           & \SI{}{\dBm}  \\ 
    Noise Figure 		& $N_{\mathrm{fig}}$ 		& \SI{6}            & \SI{}{\dB}  \\
    Noise Floor 		& $N_{\mathrm{fl}}$ & \SI{-174}       & \SI{}{\dBm}  \\  
    BI IEEE 802.11ad & & \SI{30}                   & \SI{}{\milli\second} \\ 
    DSRC beacon interval & & \SI{100}                   & \SI{}{\milli\second} \\ 
    Position update interval & & \SI{1000}                   & \SI{}{\milli\second} \\ 
	\end{tabular}
\label{tab:parameters}
\end{table}

\begin{figure}[t]     
\centering
\includegraphics[width=0.90\columnwidth]{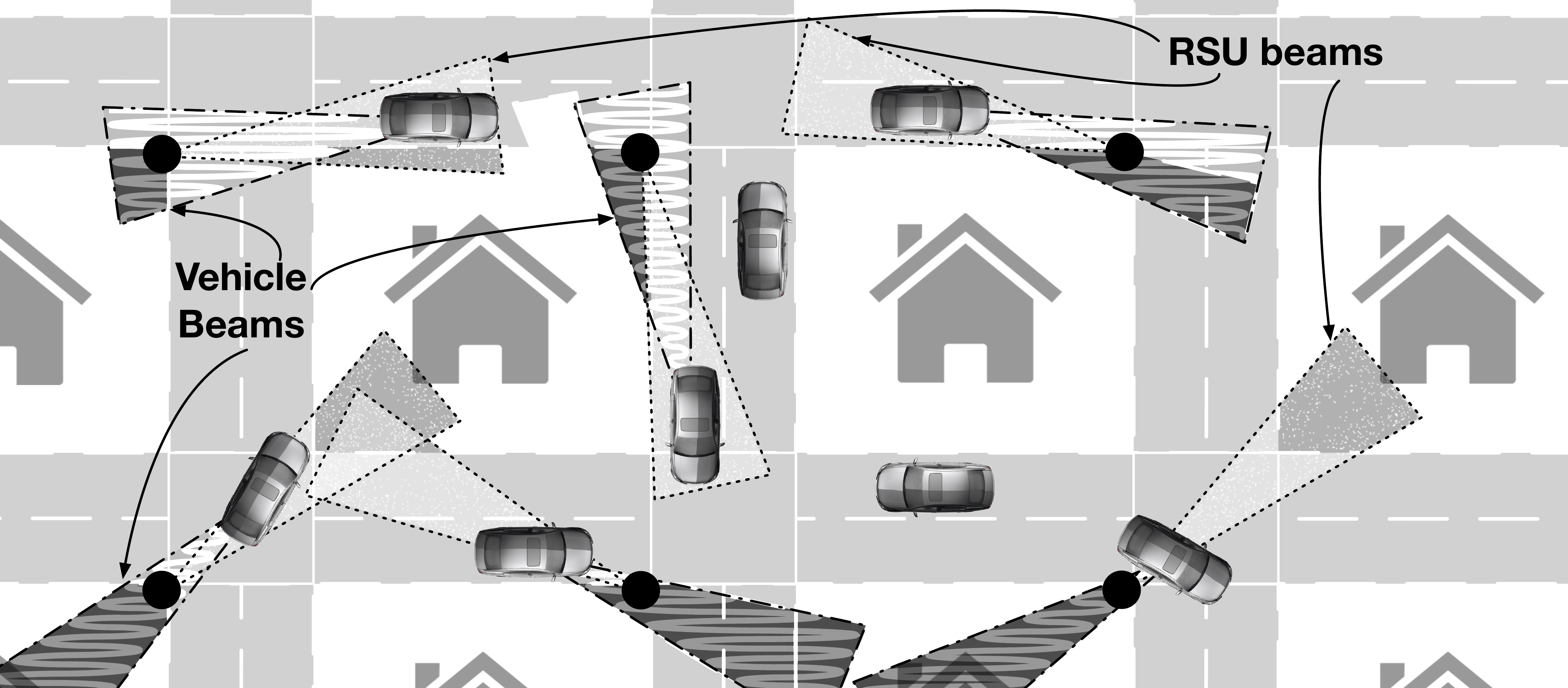}
    \caption{System level simulation scenario: Vehicles drive around a Manhattan Grid-like road network and the system performance is evaluated. The darker beam area represents the coverage region blocked by the buildings.}
    \label{fig:grid}
\end{figure}

\vspace{-0.1cm}
\subsection{Simulation Framework} \label{subsec:scenario}
As shown (Fig.~\ref{fig:coll_timemisspent}), IEEE 802.11ad performance degrades significantly as the number vehicles increases. SAMBA, having zero in-band overhead, is expected to outperform IEEE 802.11ad for a large number of vehicles. To that extent, SAMBA performance will be compared with the legacy BF technique being evaluated under various scenarios with different number of vehicles, velocities and position errors.

We utilised a \SI{200}{\meter}$\times$\SI{200}{\meter} sized Manhattan Grid road network, consisting of five horizontal and perpendicular roads (Fig.~\ref{fig:grid}). Each road is formed by four \SI{3.2}{\meter}-wide lanes (2 for each direction). The RSUs are positioned at the top-right corner of each building block on the same plane as the vehicles. The distance between two RSUs is \textasciitilde\SI{48}{\meter}. These dimensions and RSU positioning were chosen, in order to have an overlap area between the beams so the vehicles are always within the coverage region, avoiding blockages from the buildings.  The motion of the vehicle is random as described in Sec.~\ref{subsec:mob_model}. A seamless handover was assumed between the RSUs. The vehicle position error is between \SI{1}{}-\SI{3}{\meter}, following a more conservative approach compared to the centimetre accuracy presented before. RSUs positions are a priori known, so no position errors are introduced for them. Finally, when the beamwidth adaptation was not considered, the constant angle was set to $15\degree$. The rest of the simulation parameters can be found in Tab.~\ref{tab:parameters}.

\begin{figure*}[!htb]
\minipage{0.48\textwidth}
\centering
    \includegraphics[width=1\columnwidth]{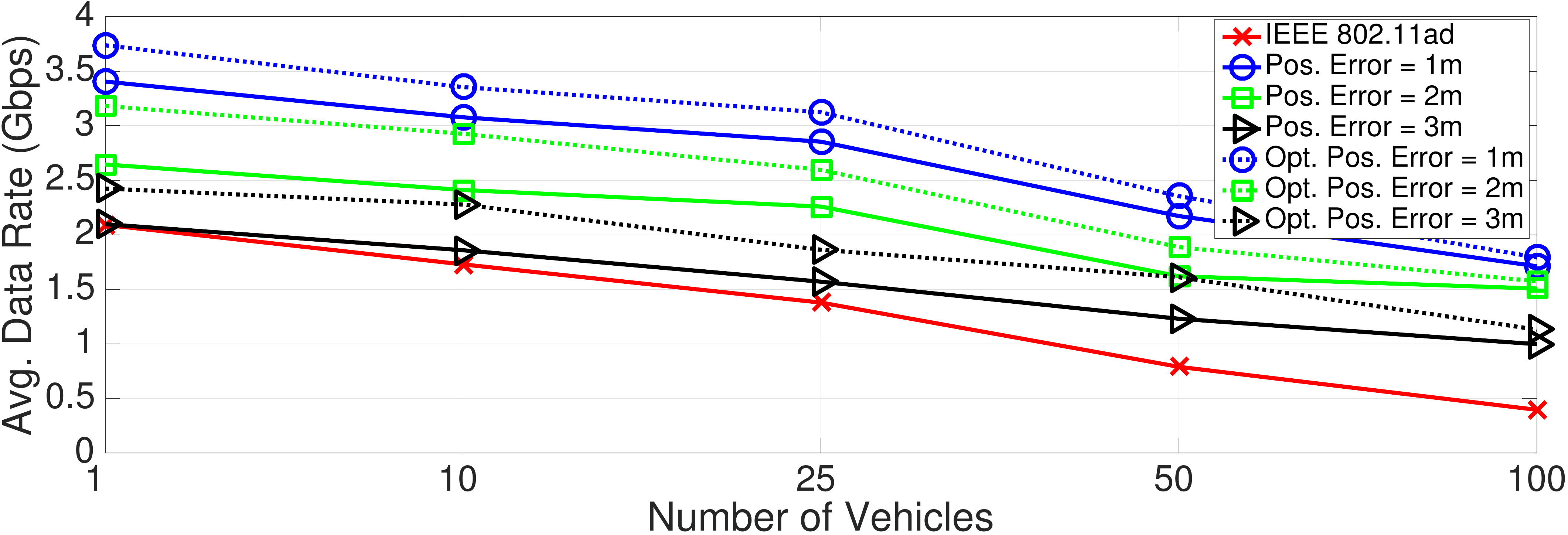}
    \caption{Average data rate per vehicle for different number of vehicles and position errors. The average speed is \SI{14}{\meter\per\second}.}
    \label{fig:many_vehicles}
\endminipage\hfill
\minipage{0.48\textwidth}
\centering
\centering
    \includegraphics[width=1\columnwidth]{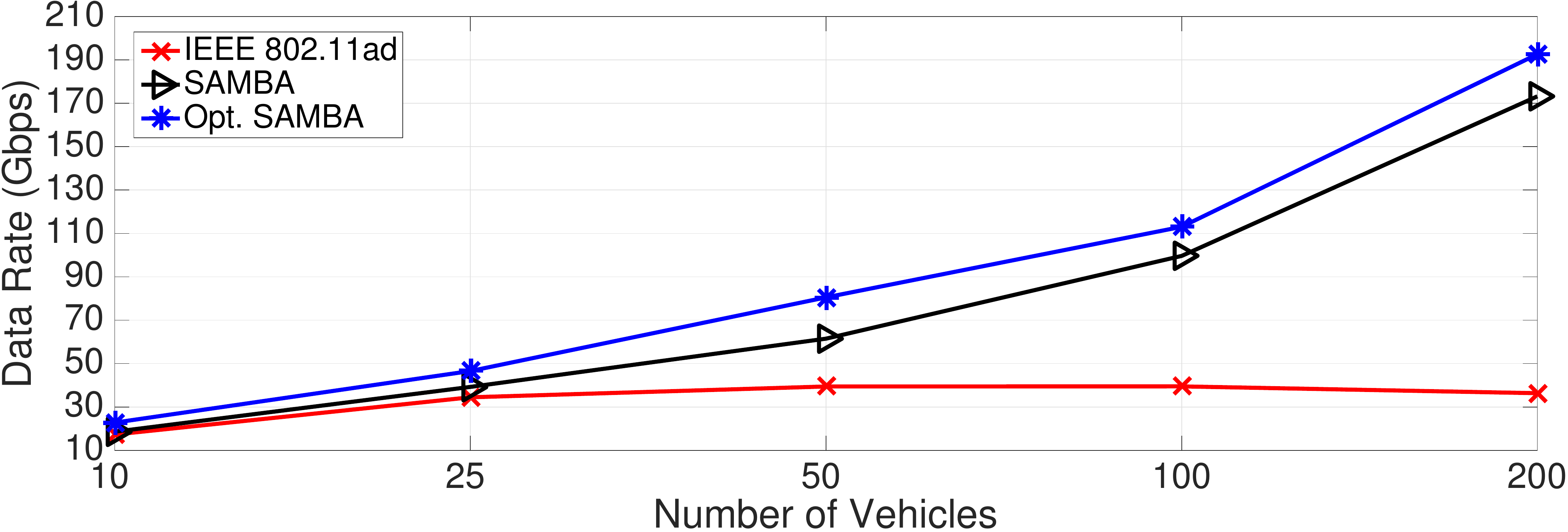}
    \caption{Throughput for the entire network. Velocity is \textasciitilde\SI{14}{\meter\per\second} and the position error for SAMBA is \SI{3}{\meter}.}
    \label{fig:network}
\endminipage\hfill
\minipage{0.48\textwidth}
\centering
    \includegraphics[width=1\columnwidth]{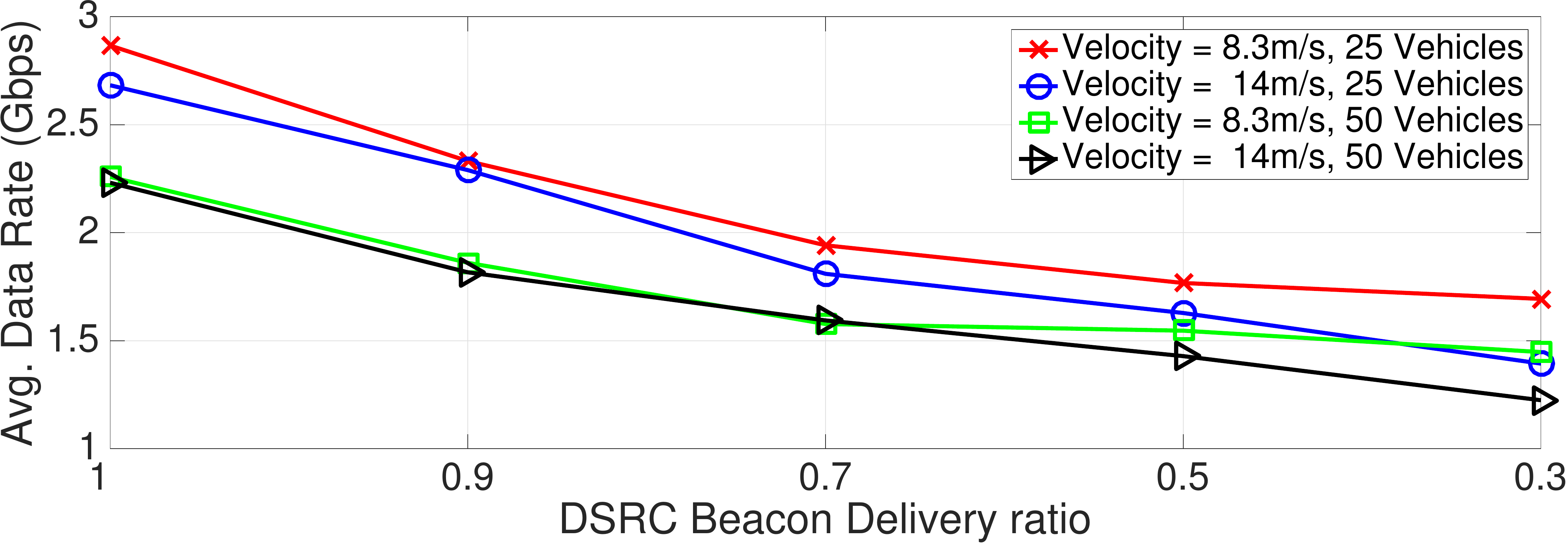}
    \caption{Average data rate per vehicle as a function of the beacon delivery ratio and the velocity. The position error was set to \SI{1}{\meter}.}
    \label{fig:prob_del}
\endminipage\hfill
\minipage{0.48\textwidth}
\centering
  \includegraphics[width=1\linewidth]{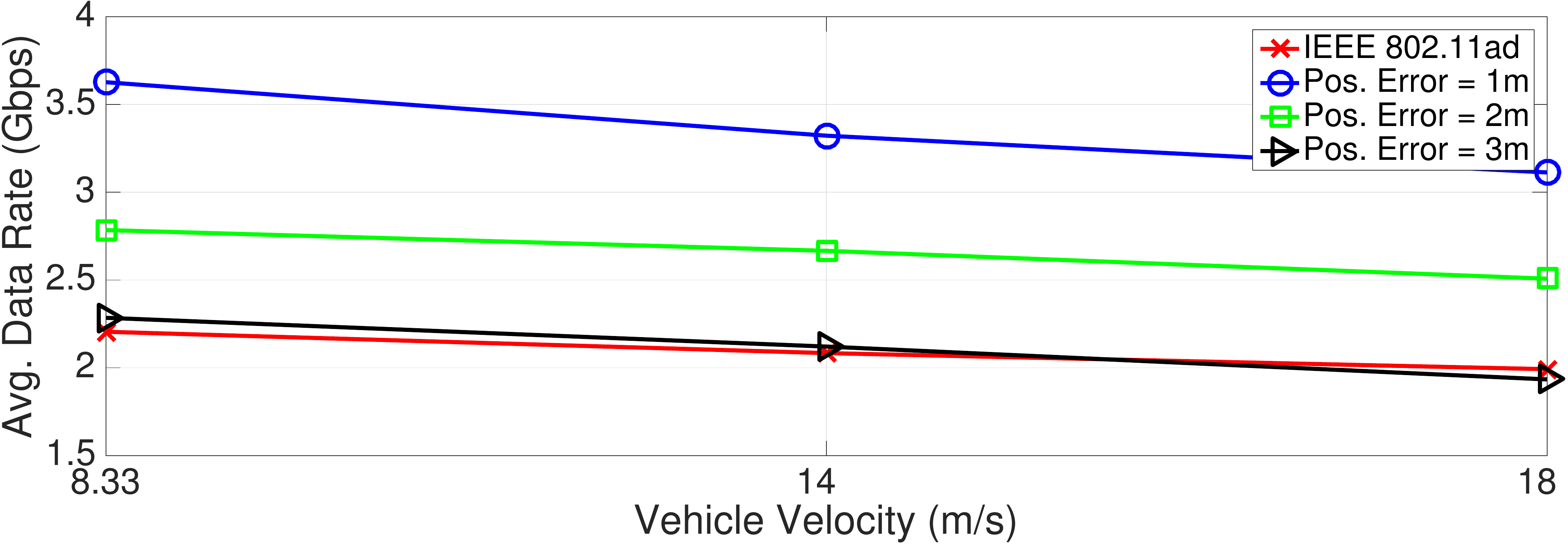}
    \caption{Achieved average data rate per vehicle for SAMBA algorithm in contrast with the legacy BF strategy for different velocities.}
    \label{fig:diff_gps_errors}
\endminipage\hfill
\end{figure*}

\vspace{-0.1cm}
\subsection{Simulation Results}
At first, the performances of SAMBA and legacy BF were evaluated for a different number of vehicles and position errors. Two different scenarios were considered for SAMBA (with and without the beamwidth adaptation). As shown in Fig.~\ref{fig:many_vehicles}, SAMBA can notably improve the system performance as it minimises the BF overhead. As expected, increasing the position accuracy improves the performance. For a position error of \SI{3}{\meter}, without beamwidth adaptation and a very sparse network ($\leq 10$ vehicles), both techniques have similar performance. However, when the vehicle density is increased, the number of collisions during A-BFT interval is increased as well (as shown in Fig.~\ref{fig:coll_prob}), significantly degrading the performance of the legacy BF. On the other hand, SAMBA can recompense with the increased density as beam alignment is based on out-of-band feedback information. 

Evaluating the throughput performance for the entire network (Fig.~\ref{fig:network}), it is observed that SAMBA significantly outperforms IEEE 802.11ad BF procedure. For dense networks the collisions during A-BFT interval limit IEEE 802.11ad performance and sightly degrade it under ultra-dense scenarios (e.g., \SI{200}{} vehicles). SAMBA, on the other hand, is able to exploit the network resources a lot better.

With respect to the previous two scenarios, and due to the physiology of the beams and the behaviour of the vehicles on a road (approach a RSU from distance, pass by it and fend off again) the beamwidth should be dynamic to always achieve the maximum data rate. Therefore, the beamwidth adaptation used in SAMBA manages to improve the performance even further (Figs.~\ref{fig:many_vehicles} and~\ref{fig:network}). Centring the vehicle within the beam manages our algorithm manages to compensate with the increased position error (e.g., \SI{3}{\meter}), maximising the SNR and consequently the data rate.

In Fig.~\ref{fig:prob_del}, SAMBA is evaluated with respect to the DSRC beacon delivery ratio. The feedback information within the DSRC beacons are the core of our system making it is obvious that the performance degrades when the beacon loss is increased. However, even with significant beacon loss ($\geq 50 \%)$ (Fig.~\ref{fig:prob_del}), SAMBA manages to achieve the required gigabit-per-second performance for future ITS services. This is because vehicles do not change their direction so often and so, macroscopically tend to move on straight line, making our algorithm tolerant to the microscopic beacon drop interval. Finally, for increased density the performance degrades due to the increased number of misalignments and consequently the waste of network resources.


Finally, evaluating SAMBA performance with respect to the velocity (Fig.~\ref{fig:diff_gps_errors}), it was shown that even though the performance slightly decreases as the velocity is increased, SAMBA can compensate well with the increased mobility.

Overall, SAMBA achieved improved system performance compared to IEEE 802.11ad. Particularly, reducing the position error will significantly improve the system performance. Also, an optimum solution for the prediction update interval is expected to enhance SAMBA performance. All in all, SAMBA was proven capable of replacing the legacy BF technique under urban vehicular scenarios.

\section{Conclusions}\label{sec:conclusions}
In this paper, an smart BF training mechanism was presented. The proposed strategy is able to achieve overhead-free BF. Exploiting feedback information broadcast over DSRC links, we introduced an agile motion-prediction model capable of estimating the position of vehicles and predicting their motion. The average data rate per vehicle as well as the network throughput were evaluated, under an urban scenario. Results showed that SAMBA outperforms the legacy sector sweep IEEE 802.11ad BF, as it can overcome beam misalignment problems and minimize the BF overhead. What is more, proposing a smart beamwidth adaptation algorithm that compensates with the vehicle movement and the beam shape, we managed to enhance the system performance even further. As such, SAMBA is a viable solution for the mmWave BF training over next-generation ITS networks. In the future, the blockage effect of the vehicles and a inter-vehicle scenario will be considered.

\bibliographystyle{IEEEtran}
\bibliography{bib.bib,IEEEabrv}
\end{document}